# Personality in Healthcare Human Robot Interaction (H-HRI): A Literature Review and Brief Critique

Connor Esterwood
School of Information
University of Michigan
Ann Arbor, Michigan, USA

Lionel P. Robert
School of Information
Robotics Institute
University of Michigan
Ann Arbor, Michigan, USA

## ABSTRACT

Robots are becoming an important way to deliver health care, and personality is vital to understanding their effectiveness. Despite this, there is a lack of a systematic overarching understanding of personality in health care human robot interaction (H-HRI). To address this, the authors conducted a review that identified 18 studies on personality in H-HRI. This paper presents the results of that systematic literature review. Insights are derived from this review regarding the methodologies, outcomes, and samples utilized. The authors of this review discuss findings across this literature while identifying several gaps worthy of attention. Overall, this paper is an important starting point in understanding personality in H-HRI.

## 1 INTRODUCTION

Robots as health care workers is one solution to the increase in the demand of health care services and the shortage of health care workers; the recent COVID-19 outbreak has only further increased their need. Even before the COVID-19 outbreak, the demand for health care services was expected to far outpace the availability of health care workers [5, 12, 33]. This was largely because of the projected increase in the population age 60+ years and older, which is expected to rise from 12.3% to 22.0% of the global population by 2050 [30]. Robots as health care workers have been introduced as a way to increase access and extend the reach of existing workers [3, 5, 12, 33]. However, during the COVID-19 pandemic, robots have been utilized to conduct health screenings, transport medical goods, and even direct patient care [24, 29, 37, 47, 53]. Taken together, it is clear that an important area of study in human–agent interaction is understanding how robots can be effective health care workers. The importance of personality in human–human health care has led HRI researchers to examine the role of personality in robot–human health care. Despite the attention, there is a lack of a systematic overarching understanding of personality in the robot–human health care literature. This makes it difficult to understand what we know and to identify what we do not know in this area. It also creates a barrier to the organization and integration of design solutions across the broader community. At present, it is difficult to know, for example, whether there is growing consensus regarding which personality traits a robot should or should not have or even if robot personality matters at all. The result is a fragmented and incoherent view of both the research area and its related design space. Nonetheless, the use of robots as health care providers is only expected to increase. This necessitates a need to reflect on what has been done in this area and to contemplate what still needs to be done. The goal of this review was to begin this process in hopes of providing insights into what we are currently doing and to help identify what we should be doing to advance the area of personality in health care HRI (H-HRI).

To accomplish this, our review offers three contributions to the literature. First, it presents the results of a systematic literature review on personality in H-HRI. Second, it derives insights into what methodologies, outcomes, and samples have been employed. Third, this review discusses findings across this literature while identifying several important gaps. These contributions are directed at informing and guiding researchers in the burgeoning field of H-HRI.

## 2 LITERATURE REVIEW

We conducted a literature review to identify the academic work related to personality in H-HRI. Next, we provide a step-by-step description of the process involved in the literature review.

*2.0.1 Search Process.* The literature search employed multiple searches via Google Scholar, the ACM Digital Library, IEEE Explore, and Scopus.

*2.0.2 Search Terms.* For this search we used six search terms. The results of these searches were manually reviewed on a search engine result page (SERP) basis using our initial inclusion criteria. SERPs were paged through progressively until no single result on the list met the specified criteria. Results prior to the page with no relevant results were extracted for review while subsequent results were not. Each SERP contained 10–25 results (depending on database) by default. In total, we found 1,819 results across all of our searches before taking into account duplicate entries.

*2.0.3 De-Duplication.* Search results were exported from Google Scholar in .bib format using the "publish or perish" application [18] and imported into r for processing. The other databases' results were exported using their respective built-in tools. De-duplication was conducted using the revtools package [44]. We identified duplicate articles on the basis of title using fuzzy matching and followed up with manual screening. Then, duplicates were removed leaving a total of unique entries numbered 1,069.

*2.0.4 Eligibility Criteria.* Given the objectives of this review, we used a three-stage approach whereby we started with the broadest of eligibility criteria and focused the review further by applying stricter criteria progressively. The initial eligibility criteria were used in the page-by-page review of search results and in all subsequent screenings. The secondary eligibility criteria were implemented in title screening and all subsequent screenings. The final eligibility criteria were applied for abstract and full-text screening. The exclusion criteria were used throughout all steps of this review.

Papers were selected for inclusion if they met three specific criteria. First, studies were required to be classified as articles or academic works that excluded patents and popular press articles. Second, studies were required to be written in the English language. The reason for excluding non-English-language publications relates to the lack of a specialist or translator on the review team making these studies difficult to screen appropriately. The third criterion for our initial eligibility was that the titles or abstracts retrieved must have explicitly mentioned both the term "robot" and the term "personality." At the secondary level, papers were selected on the basis of four additional eligibility criteria. First, studies were required to be empirical in nature and design. Second, these studies were required to use embodied physical action (EPA) robots. Third, studies were required to include measures of human or robot personality. Fourth, studies were found eligible only if they involved humans interacting with the selected EPA robots.

The final criterion used for including studies in this review required not only that the study meet all of the aforementioned eligibility requirements but also that they operated within a health care context. For the purposes of this review, a health care context was any environment where the activities and interactions performed were directly related to an individual's physical well-being. Studies were excluded if they (1) focused on embodied virtual action (EVA) (i.e. virtual agents), (2) focused on tele-presence robots, (3) focused only on manipulating robot personality without examining its impact on humans, or (4) focused only on negative attitudes toward robots (NARS) as the personality trait of interest. The exclusion of studies that used the NARS scale was based on this scale's use as a control variable in many studies (see [49], [22]).

*2.0.5 Screening Procedure.* Title screening was conducted manually in the revtools environment on the 1,069 unique entries previously identified. Screening was done only on the article title with author names and publication name hidden. Title screening was conducted based on the initial eligibility criterion. This screening identified 197 eligible studies.

Abstract screening was conducted manually in the revtools environment on the previously screened 197 studies. Abstracts were extracted from google scholar and manually added to the data-set utilized by revtools. This approach was adopted as google scholar has no native export and the exporting of abstracts on behalf of "publish or perish" is incomplete and contains missing data. This screening utilized all previous eligibility criteria in addition to the secondary eligibility criteria. After identifying 84 studies that met our secondary eligibility criteria, abstract screening was conducted a second time utilizing all previous eligibility criteria in addition to the final eligibility criteria. After this second abstract screening, 13 studies were selected for full-text screening.

In addition, 50 other potential references were identified from previously published review papers on the topic [see [22]]. All papers identified via this means were reviewed in the same way as the papers identified by our search (title and abstract screening) and with identical criteria. Ultimately, seven of the additional 50 references were found to be eligible for full-text screening.

Full-text screening involved reading each of the 20 selected papers in detail to determine their suitability based on all previously listed criteria. After completing this screening, two papers were excluded from this analysis because they reported on the same study [38]. The two excluded studies were: [39] and [40]. Figure 1 visually represents this review process and the associated counts.

## 3 REVIEW RESULTS
### 3.1 Publication Outlets

The literature review search identified 18 total papers published on the subject of human–robot interaction and personality in a health care context that met the criteria. These publications were primarily in conferences (10) and journals (7), with only one study appearing in the form of a workshop paper. A breakdown of publications by type is visible in Figure 2. In terms of specific venues, there was no dominant publication venue, with 8 of 19 studies being published in unique venues. Of the studies that were published in the same locations, four of these were published at the ACM SIGCHI conference and two were published in the International Journal of Human–Computer Studies.

Most publications were in outlets focused on human–computer interaction (5), human–robot interaction (3), interactive systems (2), and human factors, robotics, and controls engineering (3). The remaining studies ranged significantly, with two published in outlets focused on broad psychological subject matter, two published in outlets focused on aging and assistive technology, one published in an interdisciplinary open-source journal and the remaining paper published in an outlet focused on emotion, social signals, sentiment, communication. Notably, there was a lack of papers published in medicine-specific outlets. In terms of publication year, most studies were published between 2014 and 2018 as opposed to between 2002 and 2014. A breakdown of publications by year is in Figure 3.

### 3.2 Sample Data

*3.2.1 Sample Size:* The sum of all participants across studies was 805, with an average sample size of 44.7. The standard deviation for sample sizes across studies was 40.5. The standard deviation, was relatively large when compared the sample sizes across studies. However, as shown in Figure 4, a few studies had relatively large sample sizes (i.e. more than 50). Specifically, we see that the samples utilized in [7, 41] and [31] are atypical with sample sizes of 164, 114, and 98, respectively. If we exclude these samples from the calculation of mean sample size across studies we see the average sample size shrink to 23.8 with a standard deviation of 15.7. This is a fairly small average sample size and depending on the number of statistical analyses conducted can lead to a lack of power. A lack of power can result in a higher propensity of type 2 errors. Type 2 errors occur when there are significant differences but none are actually found.

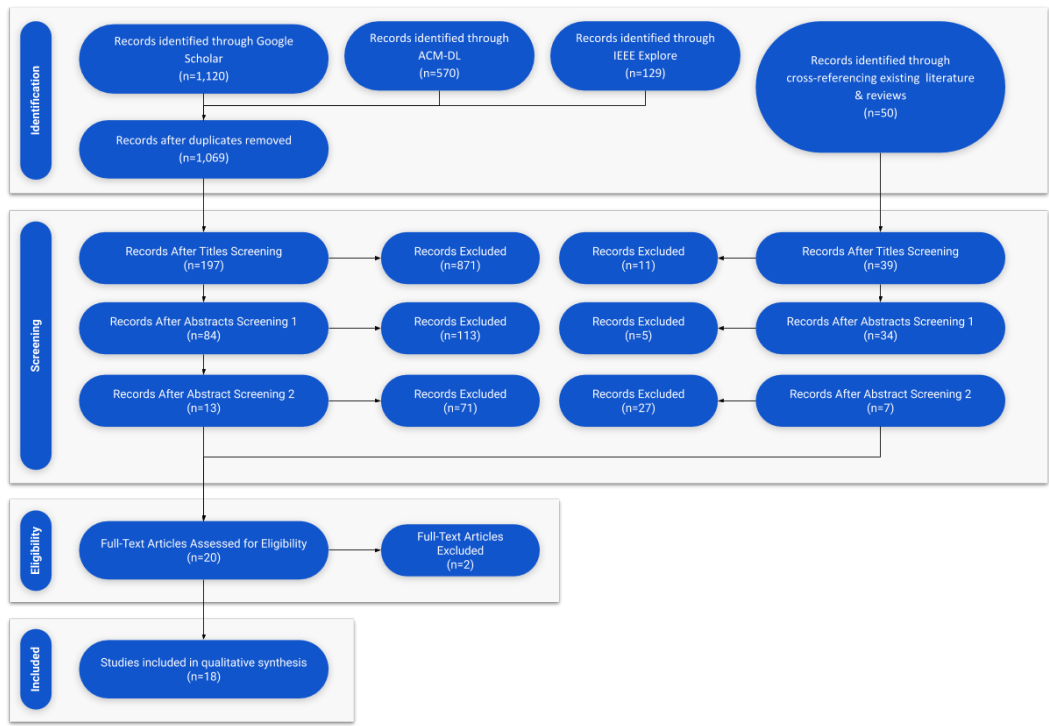

**Figure 1: Prisma Flow Diagram of Literature Review Process**

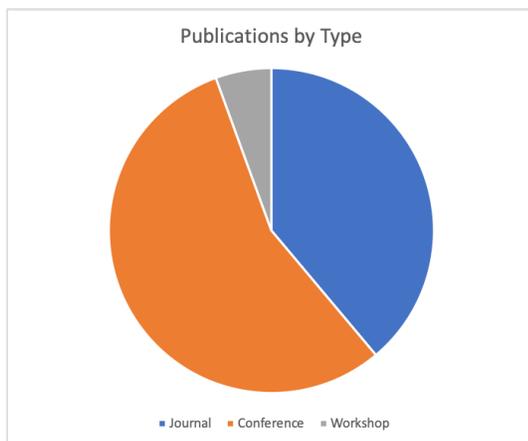

**Figure 2: Publications by Type**

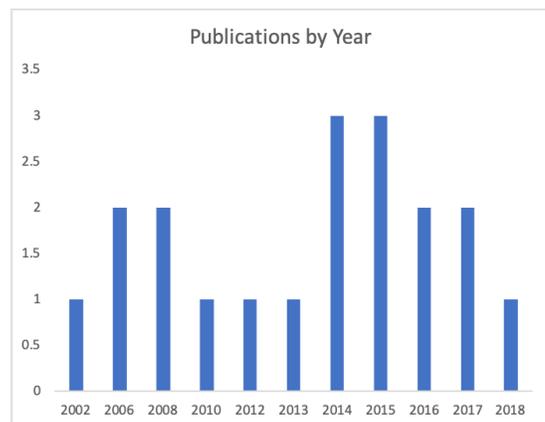

**Figure 3: Publications by Year**

*3.2.2 Participant Ages:* Age is more or less evenly distributed. The mean age across all studies was 47 and the standard deviation was 25. Specifically, six studies' average ages fell between 18 and 44 years, three fell between 45 and 64 years, and four had subjects 65 years or older. This is a fairly representative age range across studies, as is evident in Figure 5. Notably, the large number of 65+ represented in this review is the result of a handful of studies taking place in rehabilitation or retirement communities. Given the location of this sample, it is possible that the 65+ population is not encompassing of the independently living 65+ populations. Notably, four studies in this review did not report age groups [15, 31, 38, 43] and three reported only age ranges [9, 11, 14].

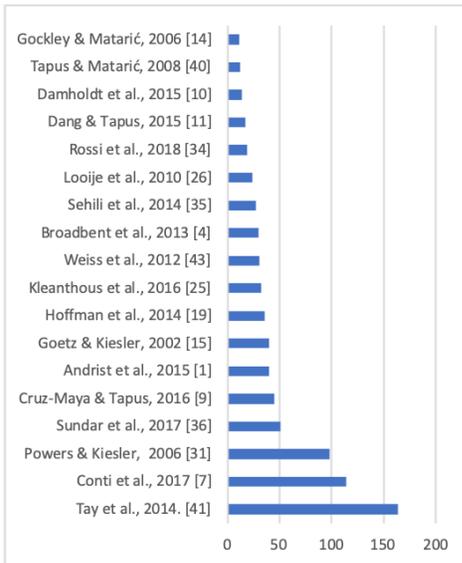

**Figure 4: Participants by Study**

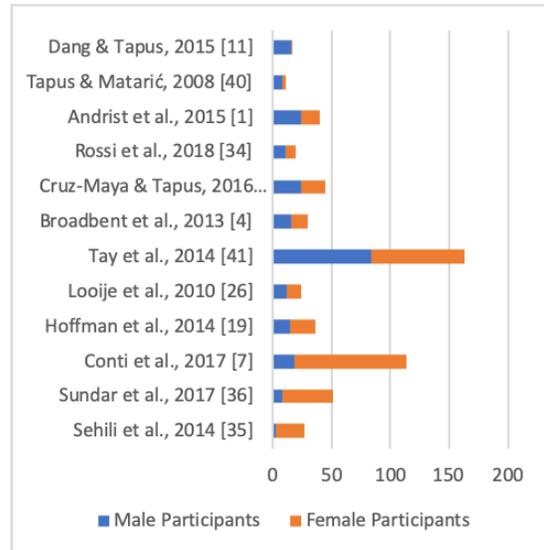

**Figure 6: Gender Balance by Study**

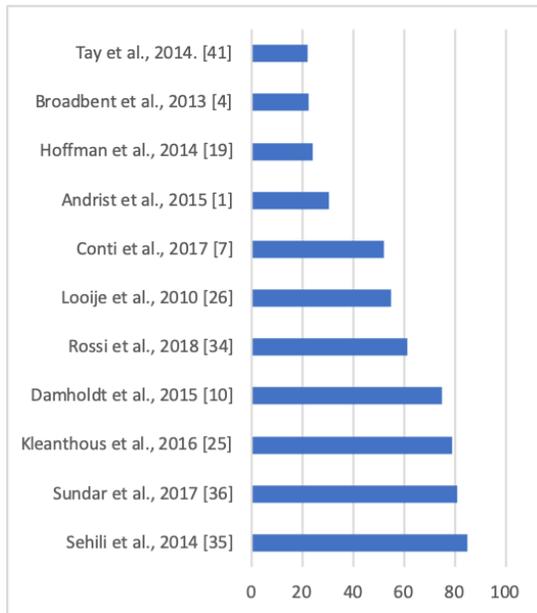

**Figure 5: Average Participant Age by Study**

*3.2.3 Gender:* We examined the distribution of gender across studies. Across all studies sampled, the percentage of women represented in this review is 57% and the percentage of men is 43%. Of all the studies, three did not provide gender information [10, 15, 31] while one only stated the majority [25]. Notably, there was significant variation among studies, with some having more than 70% male samples and others having more than 70% female samples. This creates a scenario where the average distribution of men to women seems fairly balanced overall but in individual studies this distribution was uneven. Figure 6 demonstrates this trend, with the top half of studies represented having more men and the bottom having more women in their samples. There is evidence, as a result, that within this literature gender is not being represented evenly in most studies and thus results of these studies are less generalizable than they might at first seem.

*3.2.4 Nation of Origin.* When examining the countries from which studies recruited their samples, three studies took place in more than one country while the remainder took place in the same country. Overall, samples were collected across all but one of the world's major regions. Europe was the most frequently sampled region by far, with 64% (11) of reported samples. The remaining regions represented were the Middle East and Africa 12%(2), Asia 12%(2), and North America 12%(2). No samples were found to be from South America or Central America. Notably, the majority 55% (10) of studies failed to provide region or country information in relation to their samples.

## 3.3 Level of Analysis

All studies were reviewed to determine their level of analysis. Generally, level of analysis for a particular study is at an individual, group, or organizational level. All studies included in this review were executed at the individual level, leaving the group and organizational dynamics not investigated.

## 3.4 Personality Traits

Overall, the literature examining personality in H-HRI employed five types of personality scales, which are shown in Table 1. However, the Big Five personality scale was commonly used either in whole or in part by the majority (8) of the reviewed studies. Alternative scales varied significantly among the remaining non-Big-Five studies, with two using scales either from or based on [46] two using the NEO-EFI personality inventory, and one using the Eysenck Personality Inventory (EPI). The remaining studies used different scales from one another, and notably one study [25] failed to provide details on the scale employed. Across all studies the most common dimension of personality studied was introversion vs. extraversion, which was measured in 13 of 18 studies. The studies and their associated measures are outlined in Table 1.

## 3.5 Outcomes

Outcome measures varied significantly and studies often employed more than one. In all there were 25 different outcomes. The 25 outcomes can be grouped into four broad categories: performance, acceptance, social/emotional, and robot personality. Performance accounted for 36% of the outcomes investigated. Generally, performance measures ranged from perceptual measures of how well the human or robot accomplished a task to objective measures represented by task scores or task time.

The second broad category of outcomes was classified as acceptance. This category was made up of studies that looked at acceptance, usage time, preference, trust, distance, and satisfaction. Acceptance accounted for 36% of outcomes. The third category was that of social/emotional outcomes. This category encompassed studies of attachment, cooperativeness, empathy, friendliness, warmth, social presence, and likeability. This category accounted for 21% of outcomes. The fourth category was perceptions of the robot itself. This category included measures of the human's perceptions of the robot's personality or degree of anthropomorphism. This outcome made up 7%. Table 2 provides an overview of studies and the outcomes they investigated.

## 3.6 Themes

We identified three thematic uses of personality in the H-HRI literature: human personality, robot personality, and human and robot personality.

*3.6.1 Theme 1: Human Personality.* Studies that investigated human personality exclusively represented up to 39% of the studies. These studies investigated how participants' personality impacted their interaction experience with the robot. Studies of this kind typically used participant personality as the independent variable and measured this personality via the Big Five index. For example, [7] measured human personality characteristics to determine whether different scores across the Big Five personality traits led to differences in acceptance and associated interactions with robots.

*3.6.2 Theme 2: Perceived Robot Personality.* Studies that investigated robot personality exclusively represented up to 44% of the studies. These investigated how the robot's personality impacted the human's interaction experience with the robot. For example, Goetz et al. [15] investigated how a robot's perceived personality manipulated by its interaction style (playful vs. serious) impacted the human's compliance with the robot and the robot's perceived intelligence. Notably, studies of this kind preferred to use non-Big-Five measures of personality.

*3.6.3 Theme 3: Human and Robot Personality.* Studies that investigated both human and robot personality, which represented 17% of studies, investigated how both the participant's and the robot's personalities impacted the human's interaction experience with the robot. These studies varied significantly in their aims and approaches but each sought to investigate possible interaction effects or moderators of the human's personality traits on the same human's perception of the robot's personality. For example, Looije et al., [26] investigated the relationship between a human's different personality traits and how this related to preference for one of two different personalities in a robot. These studies used various scales to measure extroversion and introversion.

## 3.7 Findings

Findings from the literature paint a clear picture that personality does indeed appear to play a role across many differing outcomes. Outcomes were fairly evenly distributed among performance (36%), acceptance (36%), and social/emotional outcomes (21%), but robot personality as an outcome was used in only a small number of studies (7%). There was also a close split between studies focusing on the theme of human personality (39%) and those focusing on the theme of robot personality (44%). Studies that investigated the theme of human and robot personalities were the minority, accounting for only three (17%) of the studies.

*3.7.1 Theme 1: Human Personality.* Studies exclusively investigating a human's personality and its relationship to HRI in a health care context made up 36% of studies in this review. Most of these studies focused on acceptance outcomes (4), performance (3), social/emotional outcomes (2), and robot personality (1).

*Performance outcomes.* Rossi et al. [34], Dang et al. [11], and Cruz-Maya et al. [9] each investigated performance. Rossi et al. [34] and Dang et al. [11] both examined extroversion and found similar results. Specifically, they found that a human's degree of extroversion did not significantly relate to measures of his or her performance. Research has found conflicting results regarding the human personality trait neuroticism / emotional control. Cruz-Maya et al. [9] found that neuroticism was positively related to human performance, whereas Rossi et al. [34] did not. Rossi et al. [34] was the only study that examined the relationships between agreeableness, conscientiousness, and openness in terms of human performance. Their results indicated that only openness had a positive relationship with human performance.

*Social/emotional outcomes.* Several studies examined the impacts of human personality on social/emotional outcomes by measuring human personalities. These studies examined a subset of human personalities: neuroticism/emotional control, extroversion/introversion, and openness.

Damholdt et al. [10] found a positive relationship between low levels of neuroticism (i.e. high emotional control) and perceptions of robot relatedness. Conti et al. [7] found a significant positive relationship between neuroticism and anxiety when interacting with a robot. Conti et al. [7] and Damholdt et al. [10] both found positive relationships between extraversion and different social–emotional outcomes such as enjoyment [7] and relatedness [10]. In terms of extraversion/introversion Conti et al. [7] and Damholdt et al. [10] both found that the higher the degrees of extraversion had a positive relationship with different social/emotional outcomes such as enjoyment and relatedness.

|   | Study | Scale | Personality Traits ||||||
|---|---|---|---|---|---|---|---|---|
|   |   |   | Extraversion | Agreeableness | Conscientiousness | Emotional Stability | Openness | Other |
| Human Personality | Conti et al., [7] | Big Five Questionnaire [6] | x | x | x | x | x |   |
|   | Looije et al., [26] | Big Five Questionnaire [42] | x | x | x | x | x |   |
|   | Sehili et al., [35] | Big Five Traits [17] | x | x | x | x | x |   |
|   | Cruz-Maya & Tapus, [9] | Big Five [16] | x | x | x | x | x |   |
|   | Dang & Tapus, [11] | Big Five [16] | x |   |   |   |   |   |
|   | Gockley & Matarić, [14] | Big Five [16] | x |   |   |   |   |   |
|   | Damholdt et al., [10] | NEO-EFI [8] | x | x | x | x | x |   |
|   | Andrist et al., [1] | Big Five [21] |   |   |   |   |   |   |
|   | Tapus & Matarić, [40] | EPI [13] |   |   |   |   |   |   |
|   | Rossi, et al. [34] | NEO-PI-3 [27] | x | x | x | x | x |   |
| Robot Personality | Goetz & Kiesler, [15] | Big-Five [20] | x | x | x | x | x |   |
|   | Looije et al., [26] | New Measure |   |   |   |   |   | x |
|   | Broadbent et al., [4] | Other |   |   |   |   |   | x |
|   | Hoffman et al., [19] | Other |   |   |   |   |   | x |
|   | Powers & Kiesler, [31] | Other |   |   |   |   |   | x |
|   | Sundar et al., [36] | Other |   |   |   |   |   | x |
|   | Tay et al., [41] | Wiggins [45] | x |   |   |   |   |   |
|   | Weiss et al., [43] | Wiggins [45] | x |   |   |   |   |   |
|   | Andrist et al., [1] | Big Five [21] | x |   |   |   |   |   |
|   | Tapus & Matarić, [40] | EPI [13] | x |   |   |   |   |   |
|   | Kleanthous et al., [25] | Not Provided | x |   |   |   |   | x |

**Table 1: Personality Traits & Scales**

| Study | Outcomes ||||
|---|---|---|---|---|
|   | Performance | Social/Emotional Outcomes | Acceptance | Perception of Robot Personality |
| Andrist et al., [1] | x |   |   |   |
| Broadbent et al., 2013 |   |   | x | x |
| Conti et al., [7] |   | x | x |   |
| Cruz-Maya & Tapus, [9] | x |   |   |   |
| Damholdt et al., [10] |   | x |   |   |
| Dang & Tapus, [11] | x |   | x |   |
| Gockley & Matarić, [14] |   |   | x |   |
| Goetz & Kiesler, [15] | x | x |   |   |
| Hoffman et al., [19] | x | x |   |   |
| Kleanthous et al., [25] |   |   | x |   |
| Looije et al., [26] |   | x | x |   |
| Kiesler & Powers, [23] | x |   |   |   |
| Rossi, et al. [34] | x |   |   |   |
| Sehili et al., [35] |   |   | x | x |
| Sundar et al., [36] | x |   | x |   |
| Tapus & Matarić, [40] |   |   | x |   |
| Tay et al., [41] | x | x | x |   |
| Weiss et al., [43] | x |   | x |   |

**Table 2: Outcomes**

Conti et al. [7] and Damholdt et al. [10] both examined openness and social emotional outcomes with conflicting results. Conti et al. [7] found that openness was positively associated with job and pleasure expressed when interacting with a robot, whereas Damholdt et al. [10] did not find a significant relationship between openness and relatedness. For the remaining personality traits of agreeableness and conscientiousness, both Damholdt et al. [10] and Conti et al. [7] investigated these traits but failed to find a significant relationship between either and social/emotional outcomes.

*Acceptance outcomes.* Four studies on the theme of human personality examined acceptance. All five of the Big-Five personality traits were examined, but only Conti et al. [7] examined agreeableness, conscientiousness, and openness. Across the different studies that investigated extroversion, Dang et al. [11] and Gockley et al. [14] found that extroversion/introversion was not significantly related to acceptance, whereas Conti et al. [7] found them to have a significant positive relationship. Researchers also found contradictory results regarding neuroticism/emotional control. Conti et al. [7] did not find a significant relationship between neuroticism/emotional control and acceptance, whereas Sehili et al. [35] did. Specifically, Sehili et al. [35] found that neuroticism was negatively related to the acceptance of a robot.

Agreeableness, conscientiousness, and openness were each investigated only by Conti et al. [7] in relation to acceptance. Findings from this study revealed non-significant results for agreeableness and conscientiousness. Openness was found to be statistically significant in terms of acceptance and to have a positive impact on acceptance of robots.

*Perceptions of the robot outcomes.* Sehili et al. [35] were the only researchers to examine the impact of human personality of the perceptions of the robot. In particular, their study focused on neuroticism/emotional control and its impact on anthropomorphic perceptions of the robot. They found a positive relationship between lower levels of neuroticism and a tendency to have a lower anthropomorphic perception of the robot.

### 3.7.2 Theme 2: Perceived Robot Personality.
Studies investigating robot personality made up 44% of studies in this review. These studies typically used non-Big-Five robot personality as outcome measures. These studies focused on performance (7) followed by acceptance (7) with a minority investigating social/emotional outcomes (3) and only one study investigating robot's perceived personality as an outcome. Given the range of personality traits examined, results varied significantly in terms of the specific personality trait of measurement. Common robot personality traits were those of anthropomorphism/human-likeness, and playfulness. These were each employed by two studies. The remainder of personality traits were unique to each study.

*Performance outcomes.* The outcome of performance was significantly impacted by a variety of robot personality traits. A total of seven studies looked at different robot personality traits and their impact on performance outcomes. In terms of anthropomorphism/ human-likeness, Powers and Kiesler [31] identified a positive and significant relationship between the more anthropomorphic or human-like robots and higher performance outcomes.

In terms of the personality trait of playfulness, both Goetz et al. [15] and Sundar et al. [36] investigated this personality trait. These authors found a significant effect of playfulness on performance outcomes. Goetz et al. [15] found that serious robots had higher performance than playful robots and Sundar et al. [36] found that this relationship was dependent on the robot's assigned role. Specifically, if the robot was assigned to an assistant role, the more playful it was, the better performance measures were, whereas a robot in a companion role who was playful led to lower performance outcomes. For the remaining personality traits investigated (e.g., responsiveness, masculinity, and robot extraversion) all but responsiveness were found to have a significant impact on performance [19, 41, 43]. Specifically, extraversion and femininity were found to have positive relationships with social/emotional outcomes [41, 43].

*Social/emotional outcomes.* Each study focused on different personality characteristics: playfulness, responsiveness, or femininity. Despite this, across all three of these studies each personality was found to have positive and significant associations with various social/emotional outcomes [15, 19, 41].

*Acceptance outcomes.* Here were found studies investigated acceptance in reference to robot personality. Each of these studies examined different robot personality characteristics. These characteristics were playfulness, femininity, friendliness, directness, and extraversion. Of these characteristics, playfulness, femininity, and extraversion were found to be significant [36, 41, 43] and to have positive associations with acceptance outcomes. Notably, Kleanthous et al [25] failed to report significance tests values but claimed to have observed a positive association between friendliness and acceptance outcomes, and a negative association between directness and acceptance outcomes.

*Perceptions of the robot outcomes.* Of studies exclusively investigating robot personality, only one study used robot personality as an outcome variable. In this study, anthropomorphism/human-likeness was assessed and found to have a significant impact on whether individuals assigned a positive personality to a robot. Specifically, this study found that the more anthropomorphic the robot, the more individuals assigned it a positive personality [4].

*3.7.3 Theme 3: Human and Robot Personality.* Three studies in this review looked at both human's and robot's perceived personalities. Two of these studies [1, 40] investigated humans's personalities in terms of introversion/extraversion, and all three used introversion/extraversion (or comparable behavior) in terms of measuring a robot's personality. Two studies looked at the same combinations of traits [1, 40] but examined different outcomes.

*Performance outcomes.* One study, Andrist et al. [1], examined both human and robot personalities in combination in terms of performance outcomes. They found that the degree of personality matching between the robot's perceived personality and the human's personality was important. Specifically, extroverts preferred extroverted robots and vice-versa for introverts.

*Social/emotional outcomes.* Once again, only one study investigated the effects of both human and robot personality in combination in terms of social/emotional outcomes. Looije et al. [26] examined the degree of robot sociability (highly sociable robots vs. non sociable) and the degree of human conscientiousness (high or low). Given sociability's relationship with extraversion vs. introversion, highly sociable robots can be considered extroverted while non-sociable robots can be considered introverted. The results of this study indicated that the degree of conscientiousness of a participant had a significant negative relationship with the sociability (extraversion) of a robot. Specifically, the more social (extroverted) robot led to less positive outcomes than the less social robot (introverted) in cases where participants rated high in conscientiousness.

*Acceptance outcomes.* Two studies examined the effects of both human and robot personalities in terms of acceptance. First, Looije et al., [26] examined the impact of a more social (extroverted) vs. a less social (introverted) robot and how this impacted humans with varying degrees of conscientiousness. Results of this study were similar to findings in terms of social/emotional outcomes where highly conscientious humans had higher degrees of acceptance of less social (introverted) robots than highly social (extroverted) robots. In the second study, Taupus et al. [40] examined the effect of matching or not matching a robot's personality (introverted vs. extroverted) with a human's personality (introverted vs extroverted) on the human's acceptance of a robot. Findings from this study were similar to those of Andrist et al. [1] in that extroverts tended to have higher acceptance ratings for extroverted robots and introverts had higher acceptance ratings of introverted robots.

## 4 DISCUSSION AND OPPORTUNITIES.

Despite the importance of personality in the health care human–robot interaction literature, there are several major gaps. Next, we present research opportunities (ROs) in the literature based on important gaps. These include research opportunities related to study samples, national biases, group-level analysis, similarity attraction and social/emotional outcomes, and moving beyond the Big Five traits. We focused on these issues because they represent several of the most salient yet addressable issues going forward.

### 4.1 RO 1: Sample

Three primary issues related to the sample were identified across the studies in the review: sample size, sampling participants ages 65+ and the wide disparity with regard to gender diversity.

*4.1.1 Size.* The vast majority of studies included fewer than 50 participants in their sample, with three studies standing apart [7, 31, 41] in having out-size samples. The mean sample size excluding these large-sample studies averaged only 23.8, making generalization of results rather limited because such small samples are prone to sampling error [2]. In that the majority of studies (83%) identified in this review had relatively small participant counts, there is an opportunity for new studies to provide additional strength to these existing findings by including additional participants and increasing their relative sample size.

*4.1.2 65+ Participants in Diverse Settings.* There is a need to examine the impacts of personality in H-HRI with participants older than 65 in settings other than assisted-living/medical-residency programs. Many individuals older than 65 live home alone and might have different challenges from those living in assisted-living/medical-residency programs. Therefore, there is a need to both identify those challenges and explore the role of personality in H-HRI. This is an unexplored area of study in personality in H-HRI.

*4.1.3 Gender.* Across the studies the issue of gender imbalance was much more problematic than it might appear, with 57% women vs. 43% men in total. However, nearly two-thirds of the studies reviewed had wider gender imbalances. This makes it difficult to generalize their findings across both populations. Additional studies are needed with properly balanced samples ensuring equal representation of men and women. In doing so, these studies would provide insights that are more generalizable across populations.

*4.1.4 Explore Other Countries.* Europe was the most frequently sampled region by far, with 64% of reported samples. The remaining regions represented were the Middle East and Northern Africa (12%), Asia (12%), and North America (12%). No samples were found from South America, Central America, or sub-Saharan Africa. Notably, the majority (55%) of studies failed to provide region or country information in relation to their samples. However, if we used the location of the authors of the papers, the breakdown appears similar, with North America (28%) increasing in size and Europe (52%) as well as the Middle East and Northern Africa (8%) decreasing in size. Asia (12%) remained consistent. Once again, we still find a lack of studies with populations from South America, Central America, or sub-Saharan Africa. That being said, we should acknowledge that our focus on English-language-only articles could in part explain the lack of studies in South America, Central America, or sub-Saharan Africa. To partly address this shortcoming, we conducted a post hoc informal review for non-English-language papers on this topic. Unfortunately, we failed to identify any additional studies. Therefore, there appears to be a gap in studies with samples from South America, Central America, and sub-Saharan Africa, or at least in English-language publications.

## 4.2 RO 2: Level of Analysis

No studies focused on personality in health care HRI investigated group-level interactions. Humans and robots in a health care context are certain to have one-on-one interactions, but these are not the only kind of interactions. For example, health care services are normally carried about by a team or group of health care workers rather than one individual. Therefore, a group-level analysis might assist in the investigation of teams and teaming between humans and robots. [48–51] The lack of investigation beyond the individual level of analysis provides an opportunity for researchers.

## 4.3 RO 3: Human and Robot Personality

At present, two studies investigated the interplay between humans' and robots' personalities in H-HRI. This stream of research is particularly important for two reasons. One, in reality both the human and the robot personalities have to be taken into consideration. Therefore, understanding the interplay between them is likely to provide important insights that can be generalized into valuable design recommendations. Two, there is a growing debate in the HRI community on whether it is better to match human and robot personality or mismatch them to achieve better interactions [22, 32, 52, 54]. Answering this question in the context of H-HRI would be valuable.

## 4.4 RO 4: Beyond Big Five

The Big Five personality traits were the most employed measures of personality used. More specifically, the comparison between the impacts of extroversion vs. introversion was by far the most widely examined relationship. However, other personality traits such as helpfulness, reliability, intelligence, and confidence have been shown to be vital to helping physicians deliver quality health care to patients [28]. Yet, these personality traits were rarely examined in the study of H-HRI. Future research must explore a more diverse set of personalities that fully represents the most important traits needed by health care workers.

## 5 CONCLUSION

Robots are becoming an important way to deliver health care across the world, and personality is vital to understanding their effectiveness. To establish what we know and identify what we do not know in this area, we conducted a review involving 1,069 articles. This review identified 18 studies that met the eligibility criteria. Specifically, we examined studies that provided the results of empirical research focused on human personality and interactions with embodied physical action robots in a healthcare context. Results of this investigation were organized into three overarching themes and gaps within these themes are highlighted. This paper is an important starting point in establishing an understanding of personality in H-HRI. Future research is needed to build on this review and expand our understanding of personality in H-HRI. Specifically, another review is needed to determine if there are any differences in the role of personality for human interactions with EPA robots versus human interactions with virtual agents/tele-presence robots in healthcare.

## REFERENCES
[1] S. Andrist, B. Mutlu, and A. Tapus. 2015. Look like me: Matching robot personality via gaze to increase motivation. *Conference on Human Factors in Computing Systems - Proceedings* 2015-April (2015), 3603–3612.
[2] Norman Blaikie. 2004. Encyclopedia of Social Science Research Methods.
[3] Robert Bogue. 2011. Robots in healthcare. *Industrial Robot: An International Journal* (2011).
[4] E. Broadbent, V. Kumar, X. Li, J. Sollers, R.Q. Stafford, B.A. MacDonald, and D.M. Wegner. 2013. Robots with Display Screens: A Robot with a More Humanlike Face Display Is Perceived To Have More Mind and a Better Personality. *PLoS ONE* 8, 8 (2013).

[5] Elizabeth Broadbent, Rebecca Stafford, and Bruce MacDonald. 2009. Acceptance of healthcare robots for the older population: Review and future directions. *International journal of social robotics* 1, 4 (2009), 319.
[6] G. V. Caprara, C. Barbaranelli, L. Borgogni, and M. Secchione. 2007. Big Five Questionnaire-2.
[7] D. Conti, E. Commodari, and S. Buono. 2017. Personality factors and acceptability of socially assistive robotics in teachers with and without specialized training for children with disability. *Life Span and Disability* 20, 2 (2017), 251–272.
[8] Paul T Costa and Robert R McCrae. 1992. Normal personality assessment in clinical practice: The NEO Personality Inventory. *Psychological assessment* 4, 1 (1992), 5.
[9] A. Cruz-Maya and A. Tapus. 2016. Teaching nutrition and healthy eating by using multimedia with a Kompai robot: Effects of stress and user's personality. *IEEE-RAS International Conference on Humanoid Robots* (2016), 644–649.
[10] M.F. Damholdt, M. Nørskov, R. Yamazaki, R. Hakli, C.V. Hansen, C. Vestergaard, and J. Seibt. 2015. Attitudinal change in elderly citizens toward social robots: The role of personality traits and beliefs about robot functionality. *Frontiers in Psychology* 6, NOV (2015).
[11] T.-H.-H. Dang and A. Tapus. 2015. Stress Game: The Role of Motivational Robotic Assistance in Reducing User's Task Stress. *International Journal of Social Robotics* 7, 2 (2015), 227–240.
[12] Inbal Deutsch, Hadas Erel, Michal Paz, Guy Hoffman, and Oren Zuckerman. 2019. Home robotic devices for older adults: Opportunities and concerns. *Computers in Human Behavior* 98 (2019), 122–133.
[13] Hans Jurgen Eysenck et al. 1968. *Eysenck personality inventory*. Educational and Industrial Testing Service San Diego.
[14] R. Gockley and M.J. Matarić. 2006. Encouraging physical therapy compliance with a hands-off mobile robot. *HRI 2006: Proceedings of the 2006 ACM Conference on Human-Robot Interaction* 2006 (2006), 150–155.
[15] J. Goetz and S. Kiesler. 2002. Cooperation with a robotic assistant. *Conference on Human Factors in Computing Systems - Proceedings* (2002), 578–579.
[16] Lewis R Goldberg. 1990. An alternative" description of personality": the big-five factor structure. *Journal of personality and social psychology* 59, 6 (1990), 1216.
[17] Samuel D Gosling, Peter J Rentfrow, and William B Swann Jr. 2003. A very brief measure of the Big-Five personality domains. *Journal of Research in personality* 37, 6 (2003), 504–528.
[18] Anne W Harzing. 2007. Publish or Perish App. https://harzing.com/resources/publish-or-perish
[19] G. Hoffman, G.E. Birnbaum, K. Vanunu, O. Sass, and H.T. Reis. 2014. Robot responsiveness to human disclosure affects social impression and appeal. *ACM/IEEE International Conference on Human-Robot Interaction* (2014), 1–7.
[20] Oliver P John, Eileen M Donahue, and Robert L Kentle. 1991. The big five inventory—versions 4a and 54.
[21] Oliver P John, Sanjay Srivastava, et al. 1999. The Big Five trait taxonomy: History, measurement, and theoretical perspectives. *Handbook of personality: Theory and research* 2, 1999 (1999), 102–138.
[22] Lionel P. Robert Jr., Rasha Alahmad, Connor Esterwood, Sangmi Kim, Sangseok You, and Qiaoning Zhang. 2020. A Review of Personality in Human–Robot Interactions. *Foundations and Trends® in Information Systems* 4, 2 (2020), 107–212.
[23] S. Kiesler, A. Powers, S.R. Fussell, and C. Torrey. 2008. Anthropomorphic interactions with a robot and robot-like agent. *Social Cognition* 26, 2 (2008), 169–181.
[24] Rainer Kimmig, René HM Verheijen, Martin Rudnicki, et al. 2020. Robot assisted surgery during the COVID-19 pandemic, especially for gynecological cancer: a statement of the Society of European Robotic Gynaecological Surgery (SERGS). *Journal of Gynecologic Oncology* 31, 3 (2020).
[25] S. Kleanthous, C. Christophorou, C. Tsiourti, C. Dantas, R. Wintjens, G. Samaras, and E. Christodoulou. 2016. Analysis of elderly users' preferences and expectations on service robot's personality, appearance and interaction. *Lecture Notes in Computer Science (including subseries Lecture Notes in Artificial Intelligence and Lecture Notes in Bioinformatics)* 9755 (2016), 35–44.
[26] R. Looije, M.A. Neerincx, and F. Cnossen. 2010. Persuasive robotic assistant for health self-management of older adults: Design and evaluation of social behaviors. *International Journal of Human Computer Studies* 68, 6 (2010), 386–397.
[27] Robert R McCrae, Paul T Costa, Jr, and Thomas A Martin. 2005. The NEO–PI–3: A more readable revised NEO personality inventory. *Journal of personality assessment* 84, 3 (2005), 261–270.
[28] Ali Mohammad Mosadeghrad. 2014. Factors influencing healthcare service quality. *International journal of health policy and management* 3, 2 (2014), 77.
[29] Robin R. Murphy, Justin Adams, and Vignesh Babu Manjunath Gandudi. 2020. Robots are playing many roles in the coronavirus crisis – and offering lessons for future disasters. https://theconversation.com/robots-are-playing-many-roles-in-the-coronavirus-crisis-and-offering-lessons-for-future-disasters-135527
[30] World Health Organization. 2016. *World health statistics 2016: monitoring health for the SDGs sustainable development goals*. World Health Organization.
[31] A. Powers and S. Kiesler. 2006. The advisor robot: Tracing people's mental model from a robot's physical attributes. *HRI 2006: Proceedings of the 2006 ACM Conference on Human-Robot Interaction* 2006 (2006), 218–225.
[32] Lionel Robert. 2018. Personality in the human robot interaction literature: A review and brief critique. In *Proceedings of the 24th Americas Conference on Information Systems, Aug.* 16–18.
[33] Hayley Robinson, Bruce MacDonald, and Elizabeth Broadbent. 2014. The role of healthcare robots for older people at home: A review. *International Journal of Social Robotics* 6, 4 (2014), 575–591.
[34] S. Rossi, G. Santangelo, M. Staffa, S. Varrasi, D. Conti, and A. Di Nuovo. 2018. Psychometric Evaluation Supported by a Social Robot: Personality Factors and Technology Acceptance. *RO-MAN 2018 - 27th IEEE International Symposium on Robot and Human Interactive Communication* (2018), 802–807.
[35] Mohamed Sehili, Fan Yang, Violaine Leynaert, and Laurence Devillers. 2014. A corpus of social interaction between nao and elderly people. In *5th International Workshop on Emotion, Social Signals, Sentiment & Linked Open Data (ES3LOD2014). LREC*.
[36] S.S. Sundar, E.H. Jung, T.F. Waddell, and K.J. Kim. 2017. Cheery companions or serious assistants? Role and demeanor congruity as predictors of robot attraction and use intentions among senior citizens. *International Journal of Human Computer Studies* 97 (2017), 88–97.
[37] Li Feng Tan and Santhosh Seetharaman. 2020. Preventing the Spread of COVID-19 to Nursing Homes: Experience from a Singapore Geriatric Centre. *Journal of the American Geriatrics Society* (2020).
[38] A. Tapus and M.J. Matarić. 2008. User personality matching with a hands-off robot for post-stroke rehabilitation therapy. *Springer Tracts in Advanced Robotics* 39 (2008), 165–175.
[39] Adriana Tapus, Cristian Tapus, and Maja J Mataric. 2006. User-robot personality matching and assistive robot behavior adaptation for. *Journal of the Robotics Society of Japan (JRSJ)* 24, 5 (2006), 14–16.
[40] Adriana Tapus, Cristian Țăpuș, and Maja J Matarić. 2008. User—robot personality matching and assistive robot behavior adaptation for post-stroke rehabilitation therapy. *Intelligent Service Robotics* 1, 2 (2008), 169.
[41] B. Tay, Y. Jung, and T. Park. 2014. When stereotypes meet robots: The double-edge sword of robot gender and personality in human-robot interaction. *Computers in Human Behavior* 38 (2014), 75–84.
[42] AJ Van Vliet. 2001. Effecten van teamsamenstelling op het moreel van uitgezonden militairen (Effects of teamcomposition on the morale of dispatched soldiers)(Rep. No. TM-01-A0404). *Soesterberg: TNO technische menskunde* (2001).
[43] A. Weiss, B. Van Dijk, and V. Evers. 2012. Knowing me knowing you: Exploring effects of culture and context on perception of robot personality. *ICIC 2012 - Proceedings of the 4th International Conference on Intercultural Collaboration* (2012), 133–136.
[44] Martin J Westgate. 2019. revtools: An R package to support article screening for evidence synthesis. *Research synthesis methods* (2019).
[45] Jerry S Wiggins. 1979. A psychological taxonomy of trait-descriptive terms: The interpersonal domain. *Journal of personality and social psychology* 37, 3 (1979), 395.
[46] Jerry S Wiggins and Ross Broughton. 1991. A geometric taxonomy of personality scales. *European Journal of Personality* 5, 5 (1991), 343–365.
[47] Guang-Zhong Yang, Bradley J Nelson, Robin R Murphy, Howie Choset, Henrik Christensen, Steven H Collins, Paolo Dario, Ken Goldberg, Koji Ikuta, Neil Jacobstein, et al. 2020. Combating COVID-19—The role of robotics in managing public health and infectious diseases.
[48] Sangseok You and Lionel Robert. 2017. Emotional attachment, performance, and viability in teams collaborating with embodied physical action (EPA) robots. *Journal of the Association for Information Systems* 19, 5 (2017), 377–407.
[49] Sangseok You and Lionel Robert. 2018. Teaming up with robots: An IMOI (inputs-mediators-outputs-inputs) framework of human-robot teamwork. *Int.J Robotic Eng(IJRE)* 2, 3 (2018).
[50] Sangseok You and Lionel Robert. 2018. Trusting robots in teams: Examining the impacts of trusting robots on team performance and satisfaction. In *Proceedings of the 52th Hawaii International Conference on System Sciences*. 8–11.
[51] Sangseok You and Lionel Peter Robert. 2019. Subgroup Formation in Human-Robot Teams. *International Conference on Information Systems* (2019).
[52] Sangseok You and Lionel P Robert Jr. 2018. Human-robot similarity and willingness to work with a robotic co-worker. In *Proceedings of the 2018 ACM/IEEE International Conference on Human-Robot Interaction*. 251–260.
[53] Zhanjing Zeng, Po-Ju Chen, and Alan A Lew. 2020. From high-touch to high-tech: COVID-19 drives robotics adoption. *Tourism Geographies* (2020), 1–11.
[54] Qiaoning Zhang, Connor Esterwood, X Jessie Yang, Lionel Robert, et al. 2019. An Automated Vehicle (AV) like Me? The Impact of Personality Similarities and Differences between Humans and AVs. *2019 AAAI Fall Symposium* (2019).